 \documentclass[final,5p,times,twocolumn]{elsarticle}

\makeatletter

\@addtoreset{equation}{section}
\makeatother

\usepackage{amsmath,amsfonts,amssymb,amsthm}
\usepackage{mathrsfs}
\usepackage{graphicx}
\usepackage{braket}
\usepackage{comment}
\usepackage{enumerate}
\usepackage{color}

\journal{Elsevier}

\begin{document}

\begin{frontmatter}

%% Title, authors and addresses

%% use the tnoteref command within \title for footnotes;
%% use the tnotetext command for the associated footnote;
%% use the fnref command within \author or \address for footnotes;
%% use the fntext command for the associated footnote;
%% use the corref command within \author for corresponding author footnotes;
%% use the cortext command for the associated footnote;
%% use the ead command for the email address,
%% and the form \ead[url] for the home page:
%%
%% \title{Title\tnoteref{label1}}
%% \tnotetext[label1]{}
%% \author{Name\corref{cor1}\fnref{label2}}
%% \ead{email address}
%% \ead[url]{home page}
%% \fntext[label2]{}
%% \cortext[cor1]{}
%% \address{Address\fnref{label3}}
%% \fntext[label3]{}

\title{Fermionic solutions of chiral Gross--Neveu and Bogoliubov--de Gennes systems \\ in nonlinear Schr\"odinger hierarchy}
%\title{Fermionic solutions of chiral Gross-Neveu and Bogoliubov-de Gennes systems\\ for gap functions belonging to the nonlinear Schr\"odinger hierarchy}
%\title{All Exact Fermionic Solutions in Nonlinear Schrodinger Hierarchy for Chiral Gross-Neveu and Bogoliubov-de Gennes systems}
%% use optional labels to link authors explicitly to addresses:
%% \author[label1,label2]{<author name>}
%% \address[label1]{<address>}
%% \address[label2]{<address>}

%\author[tokyo,keiorecs]{Daisuke~A.~Takahashi\corref{cor1}\fnref{fn1}}
\author[tokyo,keiorecs]{Daisuke~A.~Takahashi\corref{cor1}}
\ead{takahashi@vortex.c.u-tokyo.ac.jp}
\author[keiorecs,rika]{Shunji Tsuchiya}
%\ead{}
\author[keiorecs,kyoto]{Ryosuke Yoshii}
%\ead{}
\author[keiorecs,hiyoshi]{Muneto Nitta}
%\ead{}
\cortext[cor1]{Corresponding Author}
%\fntext[fn1]{Footnote environment.}
\address[tokyo]{Department of Basic Science, The University of Tokyo, Tokyo 153-8902, Japan}
\address[keiorecs]{Research and Education Center for Natural Sciences, 
Keio University, Hiyoshi 4-1-1, Yokohama, Kanagawa 223-8521, Japan}
\address[rika]{Department of Physics, Faculty of Science, Tokyo University of Science, Tokyo 162-8601, Japan}
\address[kyoto]{Yukawa Institute of Theoretical Physics, Kyoto University, Kyoto 606-8502, Japan}
\address[hiyoshi]{Department of Physics, Keio University, Hiyoshi 4-1-1, Yokohama, Kanagawa 223-8521, Japan}
\begin{abstract}
%% Text of abstract
The chiral Gross--Neveu model  
or equivalently the linearized Bogoliubov--de Gennes equation 
has been mapped to 
the nonlinear Schr\"odinger (NLS) hierarchy in 
the Ablowitz--Kaup--Newell--Segur formalism 
by Correa, Dunne and Plyushchay.
We derive the general expression for 
exact fermionic solutions 
for all gap functions in the arbitrary order of 
the NLS hierarchy.
We also find that the energy spectrum of 
the $n$-th NLS hierarchy generally has $n+1$ gaps. 
As an illustration, we present the self-consistent two-complex-kink solution with four real parameters and two fermion bound 
states. The two kinks can be placed at any position and have phase shifts.
When the two kinks are well separated, the fermion bound states are localized 
around each kink in most parameter region.
When two kinks with phase shifts close to each other are placed at distance as short as possible, the both fermion bound states have two peaks at the two kinks, \textit{i.e.,} the delocalization of the bound states occurs.
\end{abstract}

\begin{keyword}
Bogoliubov--de Gennes equation \sep Chiral Gross--Neveu model \sep Nonlinear Schr\"odinger hierarchy \sep AKNS formalism %\sep Gap equation
%% keywords here, in the form: keyword \sep keyword

%% MSC codes here, in the form: \MSC code \sep code
%% or \MSC[2008] code \sep code (2000 is the default)

\end{keyword}

\end{frontmatter}

%%
%% Start line numbering here if you want
%%
% \linenumbers

%% main text
\section{Introduction}\label{sec:intro}

The Bogoliubov--de Gennes approach \cite{DeGennes:1999} was originally formulated
for dealing with spatially inhomogeneous superconductors, that consists
of solving the Bogoliubov--de Gennes (BdG) equation and the gap equation
self-consistently. It has been widely applied from condensed matter to
high-energy physics to describe various phenomena such as solitons in
charge-density-wave systems \cite{Takayama:1980zz}, conducting polymers \cite{Brazovskii1,Brazovskii2,Brazovskii3,Heeger:1988zz}, 
incommensurate spin-density-wave states in doped {\rm Cr} \cite{Machida, Fawcett} and
fermion number fractionization in relativistic quantum field theory 
\cite{Jackiw:1975fn,Niemi:1984vz}. 
In particular, 
an important application in high-energy physics is 
to the (chiral) Gross--Neveu model \cite{Gross:1974jv} and equivalently the Nambu--Jona-Lasinio model \cite{Nambu:1961tp} that are 
known to describe dynamical chiral symmetry breaking 
and resulting dynamical mass generation. 
The original BdG system reduces to these relativistic field theories 
by linearization 
known as the Andreev approximation in the theory of superconductivity.

However, it is generally a difficult task to find analytic
self-consistent solutions of the BdG and gap equations 
when the gap function is inhomogeneous. 
For real-valued gap functions in one-dimensional systems, 
analytic solutions were completely known under 
uniform boundary conditions at spatial infinities;
a real kink 
\cite{Dashen:1975xh,Takayama:1980zz}, 
a kink-anti-kink bound state (polaron) \cite{Campbell:1981,Campbell:1981dc}, 
a three-real-kink \cite{OkunoOnodera,FeinbergPLB} 
and more general real solutions \cite{Feinberg:2003qz}. 
Under a periodic boundary condition, 
a real kink crystal known as the Larkin--Ovchinnikov state 
is known \cite{larkin:1964zz,Machida:1984zz}. 
On the other hand, 
few analytic solutions in complex condensates are known;  
a twisted (or complex) kink \cite{Shei:1976mn}, {\it i.e.}, 
a kink with a twisted boundary condition on the phase. 

Recently, a breakthrough has been made by 
Ba\c{s}ar and Dunne \cite{Basar:2008im,Basar:2008ki} that a suitable ansatz 
for the Gorkov resolvent reduces the gap equation to the nonlinear 
Schr\"odinger (NLS) equation, which is exactly soluble. 
Since the derived 
NLS equation is a closed equation for the order parameter $\Delta(x)$, this
approach enables one to avoid the self-consistent calculation of the
coupled BdG and gap equations. They also found a new exact 
self-consistent crystalline condensate, 
a complex kink crystal with both phase and amplitude modulating 
(the Larkin--Ovchinnikov--Fulde--Ferrel state 
\cite{larkin:1964zz,Fulde:1964zz}), 
that includes all previously known solutions as special cases. 
This approach by Ba\c{s}ar and
Dunne has been extended to incorporate spin imbalance effect which plays
an important role in superconductors under a strong magnetic field and
spin polarized ultracold Fermi gases \cite{Yoshii:2011yt}. 

The Ba\c{s}ar--Dunne approach has been further developed by
Correa, Dunne, and Plyushchay \cite{Correa:2009xa} by introducing 
the integrable nonlinear equations for the gap function $\Delta(x)$ that
belong to the NLS hierarchy of the celebrated Ablowitz--Kaup--Newell--Segur (AKNS) formalism.
However, general expression for fermionic solutions of 
the BdG equation has not been obtained for those $\Delta(x)$; 
One had to solve the BdG equation to obtain fermionic solutions for
each given order parameter $\Delta(x)$.

The aim of the present Letter is to derive the general expression for exact fermionic solutions of the BdG
equation when the gap function $\Delta(x)$ obeys the NLS equations in the arbitrary order of the NLS hierarchy.
We first formulate the BdG system in the AKNS form and briefly review
the derivation of higher order NLS equations.
We then discuss the derivation of the general expression for fermionic
solutions in detail. 
As a concrete example, we apply our formalism to 
the two-complex-kink solution
for the second nonlinear equation (AKNS$_2$) to derive the fermionic
solutions for the bound states as well as for scattering states.

\section{AKNS form and solutions of linearized BdG equation}
\subsection{Problem considered in this Letter}
	\indent First, let us clarify the problem treated in this Letter mathematically. What we want to solve is the linearized (Andreev-approximated) BdG equation
	\begin{align}
		\begin{pmatrix} -\mathrm{i}\partial_x & \Delta \\ \Delta^* & \mathrm{i}\partial_x \end{pmatrix}\begin{pmatrix}u \\ v \end{pmatrix}=E \begin{pmatrix} u \\ v \end{pmatrix} \label{eq:BdGintro}
	\end{align}
	with the assumption that the gap  $ \Delta(x) $ satisfies stationary higher order NLS equations \cite{FaddeevTakhtajan}, known as $ \text{AKNS}_{n} \ (n=1,2,\dots) $\cite{Correa:2009xa}: 
	\begin{align}
		\mathrm{i}\Delta_t=\sum_{j=1}^{n+2}c_j (-\mathrm{i}M_{12}^{(j)}), \label{eq:HNLSintro}
	\end{align}
	where  $ c_j $s are arbitrary real coefficients, and $ M_{12}^{(j)} $s are given by
	\begin{align}
		-\mathrm{i}M_{12}^{(1)}&=\Delta, \\
		-\mathrm{i}M_{12}^{(2)}&=-\mathrm{i}\Delta_x, \\
		-\mathrm{i}M_{12}^{(3)}&=-\Delta_{xx}+2|\Delta|^2\Delta, \\
		-\mathrm{i}M_{12}^{(4)}&=\mathrm{i}(\Delta_{xxx}-6|\Delta|^2\Delta_x), \\
		-\mathrm{i}M_{12}^{(5)}&=\Delta_{xxxx}-2(|\Delta|^2)_{xx}\Delta-3\Delta^*(\Delta^2)_{xx}+6|\Delta|^4\Delta, \\
		&\vdots \nonumber
	\end{align}
	Here the subscripts $t$ and $x$ denote the differentiation with respect to $t$ and $x$, respectively. Since we consider a stationary (time-independent) problem, henceforth L.H.S. of Eq. (\ref{eq:HNLSintro}) is set to zero. How to generate $ M_{12}^{(j)} $s is reviewed in Section~\ref{sec:HNLS}. For later convenience, we also prepare the term ``\textit{pure}  $ \textit{AKNS}_n $'' for the  $ \text{AKNS}_n $ with $ c_1=\dotsb=c_{n+1}=0 $ and $ c_{n+2}=1 $.\\
	\indent The most familiar NLS equation is equivalent to $ \text{AKNS}_1 $. 
$ \text{AKNS}_2 $ is also known as the Hirota equation \cite{Hirota1973} and used to describe the soliton propagation phenomena in optical fibers. (See also Ref. \cite{SasaSatsumaJPSJ1991}.) We also note that pure $ \text{AKNS}_2$ is equivalent to the modified KdV equation when $ \Delta $ is real-valued.

\subsection{Overdetermined system and Compatibility condition}\label{subsec:compati}
	The AKNS system \cite{AKNS1974} is formulated through the following overdetermined system:
	\begin{align}
		\frac{\partial }{\partial x}\begin{pmatrix}u \\ v \end{pmatrix} = U(x,t,\lambda) \begin{pmatrix}u \\ v \end{pmatrix}, \label{eq:AKNSU} \\
		\frac{\partial }{\partial t}\begin{pmatrix}u \\ v \end{pmatrix} = V(x,t,\lambda) \begin{pmatrix}u \\ v \end{pmatrix}, \label{eq:AKNSV}
	\end{align}
	with  $ 2\times2 $  matrices $U$ and $V$ defined below, 
	and the nonlinear equation that we want to solve arises as a compatibility condition (or a zero-curvature condition)
	\begin{align}
		U_t-V_x+[U,V]=0. \label{eq:zcc}
	\end{align}
	In the case of NLS hierarchy, $ U $ is a $ 2\times 2 $ matrix defined by
	\begin{align}
		U(x,t,\lambda) &= \begin{pmatrix} -\mathrm{i}\lambda  & q(x,t) \\ r(x,t) & \mathrm{i}\lambda \end{pmatrix}. \label{eq:AKNSUform}
	\end{align}
	Here $ \lambda $ is an $ (x,t) $-independent spectral parameter. Henceforth we often omit arguments of functions when it is clear from the context. If we identify $ r, q,  $ and $ \lambda $ as
	\begin{align}
		q=-\mathrm{i}\Delta,\quad r=\mathrm{i}\Delta^*,\quad \text{and \quad } \lambda=-E,
	\end{align}
	then Eq. (\ref{eq:AKNSU}) readily reduces to the BdG equation (\ref{eq:BdGintro}). The form of $ V $ for higher order NLS equations will be given later.\\
	\indent Note that if we consider a time-independent problem, the zero-curvature condition has the same form with the Eilenberger equation:
	\begin{align}
		V_x=[U,V]. \label{eq:zerocurvaturest}
	\end{align}
	Therefore, $ V=R\sigma_3 $ immediately gives a particular solution for the Eilenberger equation up to normalization, where $ R $ is the Gorkov resolvent. (Note that $ R $ must be normalized as $ \det R = -1/4 $ \cite{Basar:2008im,Basar:2008ki}.)\\
	\indent Even though $ V $ is an $ x$-dependent matrix, the eigenvalues of $ V $ is independent of $x$, since
	\begin{align}
		(\operatorname{tr}V)_x=(\operatorname{tr}V^2)_x=0
	\end{align}
	follows from Eq.~(\ref{eq:zerocurvaturest}), and therefore a characteristic polynomial for $ V $
	\begin{align}
	\begin{split}
		\det\left( V-\mathrm{i\omega}\boldsymbol{1}_2 \right)&=\det V-\mathrm{i}\omega\operatorname{tr}V-\omega^2 \\
		&= \frac{1}{2}(\operatorname{tr}V^2-(\operatorname{tr}V)^2)-\mathrm{i}\omega\operatorname{tr}V-\omega^2
	\end{split}
	\end{align}
	becomes independent of $x$. It is worthy to note that Eq.~(\ref{eq:zerocurvaturest}) is identical to the famous Lax equation \cite{Lax} if we replace $ x $ by $ t $, and the above proof is the same as the proof of the isospectral property of the Lax operator. 
\subsection{Higher order NLS}\label{sec:HNLS}
	\indent In this subsection we review how to generate higher order NLS equations. Although the content in this subsection is only a revisit of the famous  textbook by Faddeev and Takhtajan \cite{FaddeevTakhtajan}, we briefly review it  for self-containedness  since the definition of spectral parameter in our Letter differs from theirs in twice factor.\\
	\indent The NLS equation has infinitely many conservation laws, and correspondingly, there are infinitely many integrable equations. In accordance with Ref. \cite{Correa:2009xa}, we call the equation generated from the $ n $-th order conserved quantity $ \text{AKNS}_{n-2} $.  Let $ V^{(n)} $ be a matrix which yields pure $ \text{AKNS}_{n-2} $. Then $ V^{(n)} $ can be calculated as follows\cite{FaddeevTakhtajan}. Let $ M $ be a $ 2\times 2 $ matrix which satisfies the equation
	\begin{align}
		M_x &= [U,M] \label{eq:Mdef}
	\end{align}
	and has the following formal Laurent series 
	\begin{align}
		M = \sum_{n=0}^\infty\frac{M^{(n)}}{(-2\lambda)^n}, 
\label{eq:Mexpand}  \quad
		M^{(0)} := -\frac{\mathrm{i}\sigma_3}{2}.
	\end{align}
	Substituting Eq.~(\ref{eq:Mexpand}) into Eq.~(\ref{eq:Mdef}), and using
	\begin{align}
		U = 2\lambda M^{(0)}+U^{(1)}, \quad U^{(1)}:=\begin{pmatrix}0&q \\ r&0 \end{pmatrix},
	\end{align}
	one obtains the recurrence relation
	\begin{align}
		M_x^{(n)}=\frac{\mathrm{i}}{2}[\sigma_3,M^{(n+1)}]+[U^{(1)},M^{(n)}] \quad (n=0,1,\dots). \label{eq:Mrecur}
	\end{align}
	Thus one can determine $M^{(n)}$ inductively. If one sets all arising integration constants to zero, one obtains
	\begin{align}
		M^{(1)}&=-U^{(1)}=\begin{pmatrix}0& -q \\ -r & 0 \end{pmatrix}, \\
		M^{(2)}&= \begin{pmatrix} -\mathrm{i}rq & \mathrm{i}q_x \\ -\mathrm{i}r_x & \mathrm{i}rq \end{pmatrix}, \\
		M^{(3)}&= \begin{pmatrix} qr_x-rq_x & q_{xx}-2rq^2 \\ r_{xx}-2r^2q & -(qr_x-rq_x) \end{pmatrix}, \\
		M^{(4)}&= \begin{pmatrix} \ \begin{matrix}\mathrm{i}(rq_{xx}+qr_{xx}\qquad\quad \\ \qquad-r_xq_x-3r^2q^2)\end{matrix} & -\mathrm{i}(q_{xxx}-6rqq_x) \\ \mathrm{i}(r_{xxx}-6qrr_x)  &  \begin{matrix}-\mathrm{i}(rq_{xx}+qr_{xx}\qquad\quad \\ \qquad-r_xq_x-3r^2q^2)\end{matrix} \end{pmatrix},\!\! \\
		&\vdots \nonumber
	\end{align}
	$ V^{(n)} $ for pure $ \text{AKNS}_{n-2} $  is then given by
	\begin{align}
		V^{(n)} &= \sum_{k=0}^{n-1}(-2\lambda)^{n-1-k}M^{(k)}. \label{eq:Vexpand}
	\end{align}
	Setting $ V=V^{(n)} $ in zero-curvature condition (\ref{eq:zcc}) and using (\ref{eq:Mrecur}), one obtains the higher order NLS equation
	\begin{align}
		U^{(1)}_t = \frac{\mathrm{i}}{2}[\sigma_3,M^{(n)}] \quad\leftrightarrow\quad \begin{cases} \mathrm{i}q_t = -M^{(n)}_{12}, \\ \mathrm{i}r_t=M^{(n)}_{21}. \end{cases}
	\end{align}
	Thus, the off-diagonal element of $ M^{(n)} $ gives the higher order NLS equation itself. The matrix $ V $ which yields  general (non-pure) $ \text{AKNS}_n $ is simply the real-coefficient linear combination of  $ V^{(j)} $s: 
	\begin{align}
		V = \sum_{j=1}^{n+2}c_j V^{(j)}, \label{eq:AKNSVform}
	\end{align}
	and the resultant equation is given by $ \mathrm{i}q_t = \sum_{j=1}^{n+2}c_j (-M_{12}^{(j)}) $, as shown in Eq.~(\ref{eq:HNLSintro}) with recalling $ q=-\mathrm{i}\Delta $.
\subsection{Solutions for stationary AKNS equation}
	In this subsection, we solve Eq.~(\ref{eq:AKNSU}) in static case,  namely the ordinary differential equation 
	\begin{align}
		\frac{\partial }{\partial x}\begin{pmatrix}u \\ v \end{pmatrix} = U(x,\lambda) \begin{pmatrix}u \\ v \end{pmatrix} \label{eq:AKNSU2}
	\end{align}
	under the assumption that there exists another matrix $ V(x,\lambda) $ 
satisfying the differential equation
	\begin{align}
		V_x=[U,V]. \label{eq:compatible2}
	\end{align}
	Although our main interest in this Letter is the case where $ U $ and $ V $ are given by Eqs. (\ref{eq:AKNSUform}) and (\ref{eq:AKNSVform}), the solution given in this subsection does not depend on the special form of  $ U $ and $ V $.\\
	\indent The solution is constructed from the following ansatz:
	\begin{align}
		\mathrm{i}\omega \begin{pmatrix}u \\ v \end{pmatrix}=V(x,\lambda)\begin{pmatrix}u \\ v \end{pmatrix}. \label{eq:AKNSV2}
	\end{align}
	The ansatz (\ref{eq:AKNSV2}) and the original equations (\ref{eq:AKNSU2}) and (\ref{eq:compatible2}) are ``compatible'', since $ \mathrm{i}\omega \times \text{(Eq.~(\ref{eq:AKNSU2}))} $ and $ \partial_x(\text{Eq.~(\ref{eq:AKNSV2})}) $ yield
	\begin{align}
		\mathrm{i}\omega\frac{\partial }{\partial x}\begin{pmatrix}u \\ v \end{pmatrix}=UV\begin{pmatrix}u \\ v \end{pmatrix} ,
\quad
		\mathrm{i}\omega\frac{\partial }{\partial x}\begin{pmatrix}u \\ v \end{pmatrix}=(V_x+VU)\begin{pmatrix}u \\ v \end{pmatrix},
	\end{align}
	respectively, while R.H.S. of these two are equal because of Eq.~(\ref{eq:compatible2}).\\
	\indent Even though we can treat the problem in the above framework for general $ U $ and $ V $, we restrict our discussion to the case where $ U $ and $ V $ are traceless:
	\begin{align}
		\operatorname{tr}U=\operatorname{tr}V=0,
	\end{align}
	since the final expression for $ (u,v) $ becomes a bit simpler. We note that the matrices  $ U $ and $ V $ for the NLS hierarchy, Eqs. (\ref{eq:AKNSUform}) and (\ref{eq:AKNSVform}) satisfy this condition.\\
	\indent For later reference, we write down the time-independent zero-curvature condition (\ref{eq:compatible2}) for each component:
	\begin{align}
		V_{11x}&=-V_{22x}=U_{12}V_{21}-U_{21}V_{12},\label{eq:compati01} \\
		V_{12x}&=2U_{11}V_{12}-2U_{12}V_{11},\label{compati02} \\
		V_{21x}&=-2U_{11}V_{21}+2U_{21}V_{11}. 
	\end{align}
	Here the traceless assumption is used.\\
	\indent In order for Eq. (\ref{eq:AKNSV2}) to have a non-vanishing solution,
	\begin{align}
		& \det(V-\mathrm{i}\omega \boldsymbol{1}_2)=0 % \nonumber \\
		\leftrightarrow\quad \omega^2=\det V = -V_{11}^2-V_{12}V_{21} \label{eq:Vomega0}
	\end{align}
	must hold. As discussed in Section~\ref{subsec:compati},  $ \det V $ is  independent of $x$, and $ \omega $ is determined as an  $ x $-independent constant:
	\begin{align}
		\omega = \pm\sqrt{\det V}.
	\end{align}
	From the expression (\ref{eq:Vomega0}) we also obtain an important factorization:
	\begin{align}
		V_{12}V_{21}=(\mathrm{i}V_{11}+\omega)(\mathrm{i}V_{11}-\omega). \label{eq:factorV12V21}
	\end{align}
	This relation technically plays an important role in the next subsection.
When  $ \omega $ is a root of Eq. (\ref{eq:Vomega0}), 
	\begin{align}
		\frac{v}{u} = -\frac{\mathrm{i}V_{11}+\omega}{\mathrm{i}V_{12}}=\frac{\mathrm{i}V_{21}}{\mathrm{i}V_{11}-\omega} \label{eq:uvratio}
	\end{align}
	holds from Eq. (\ref{eq:AKNSV2}). Using this, one can eliminate either of $ u $ or $ v $. If one chooses to eliminate $ v $, from the first row of Eq. (\ref{eq:AKNSU2}), one obtains a first order differential equation for $ u $ :
	\begin{align}
		u_x &= U_{11}u+U_{12}v = \left[U_{11}-U_{12}\frac{\mathrm{i}V_{11}+\omega}{\mathrm{i}V_{12}}\right]u.
	\end{align}
	With the use of Eqs. (\ref{compati02}) and (\ref{eq:factorV12V21}), one can show
	\begin{align}
		2\frac{u_x}{u} &= \frac{V_{12x}}{V_{12}}+\frac{2\mathrm{i}\omega U_{12}V_{21}}{(\mathrm{i}V_{11}+\omega)(\mathrm{i}V_{11}-\omega)}.
	\end{align}
	Furthermore, since it follows that
	\begin{align}
		U_{12}V_{21} = \frac{U_{12}V_{21}+U_{21}V_{12}}{2}+\frac{V_{11x}}{2}
	\end{align}
	from Eq. (\ref{eq:compati01}), one obtains
	\begin{align}
		2\frac{u_x}{u} = \frac{V_{12x}}{V_{12}}+\frac{\mathrm{i}\omega V_{11x}}{(\mathrm{i}V_{11}+\omega)(\mathrm{i}V_{11}-\omega)}+\mathrm{i}\omega\left( \frac{U_{12}}{V_{12}}+\frac{U_{21}}{V_{21}} \right).
	\end{align}
	Except for the rightmost term, each term can be integrated symbolically, and the solution is given by
	\begin{align}
		u^2 &= C V_{12}\sqrt{\frac{\mathrm{i}V_{11}-\omega}{\mathrm{i}V_{11}+\omega}}\exp\left[ \mathrm{i}\omega\int_0^x\mathrm{d}x\left( \frac{U_{12}}{V_{12}}+\frac{U_{21}}{V_{21}} \right) \right], \label{eq:solutemp} \\
	\intertext{and using Eq. (\ref{eq:uvratio}), the expression for $ v $ is also obtained:}
		v^2 &= -C V_{21}\sqrt{\frac{\mathrm{i}V_{11}+\omega}{\mathrm{i}V_{11}-\omega}}\exp\left[ \mathrm{i}\omega\int_0^x\mathrm{d}x\left( \frac{U_{12}}{V_{12}}+\frac{U_{21}}{V_{21}} \right) \right]. \label{eq:solvtemp}
	\end{align}
	Here $ C $ is an integration constant.  Note that there are two possible values for $ \omega $ since Eq. (\ref{eq:Vomega0}) is quadratic, and it corresponds to two linearly-independent solutions of the original equation (\ref{eq:AKNSU2}). (The case $ \omega=0 $ is rather exceptional and the second solution must be constructed by reduction of order.) Needless to say, when one takes the square root of (\ref{eq:solutemp}) and (\ref{eq:solvtemp}), one must choose the sign so that the relation (\ref{eq:uvratio}) is satisfied.\\
	\indent The solutions (\ref{eq:solutemp}) and (\ref{eq:solvtemp}) are one of the main results of this Letter. Since this expression is valid for arbitrary spectral parameter $ \lambda $, it means that the linearized BdG equation has been solved for arbitrary energy $ E $ regardless of whether the wave function of the energy $E$ diverges or not. In the next subsection, we give a criterion for a given energy $ E $ to belong to the spectrum.
\subsection{Energy spectra}
	In the preceding subsection, we have derived the general solution for the stationary AKNS equation. Here we concentrate on the specialized issues for the linearized BdG equation, i.e., we consider the case where the matrices  $ U $ and $ V $ are given by Eqs. (\ref{eq:AKNSUform}) and (\ref{eq:AKNSVform}).\\
	\indent Henceforth we always assume $ q=-\mathrm{i}\Delta $ and $ r=\mathrm{i}\Delta^* $, so $ r=q^* $.  As seen in Section~\ref{sec:HNLS},  $ V $ of NLS hierarchy always satisfies:
	\begin{enumerate}[(i)]
		\item If $ \lambda $ is real ( $ \leftrightarrow E=-\lambda $ is real), $ \mathrm{i}V_{11} $ is real.
		\item If $ \lambda $ is real,  $ V_{21}^*=V_{12}^{} $. 
	\end{enumerate}
	From these facts, the left hand side of (\ref{eq:factorV12V21}) is real and non-negative when $ \lambda $ is real.  We can further see the following:
	\begin{enumerate}[(i)]
	\setcounter{enumi}{2}
		\item If $ \lambda $ is real,  $ \det V $ is also real. Therefore,  $ \omega=\pm \sqrt{\det V} $ is either real or pure imaginary.
		\item If $ \lambda $ and $ \omega $ are real,  $ \mathrm{i}V_{11}+\omega $ and $ \mathrm{i}V_{11}-\omega $ are both real and have the same sign. Since $ \mathrm{i}V_{11}-|\omega|\le \mathrm{i}V_{11}\le \mathrm{i}V_{11}+|\omega| $,  $ \mathrm{i}V_{11} $ also has the same sign. Furthermore, if  $ \omega\ne0 $ and the gap function $ \Delta(x) $ is bounded and  $ C^k $ with $ k $ being sufficiently large,  $ \operatorname{sgn}(\mathrm{i}V_{11}) $ is globally constant and does not depend on $ x $. 
		\item As a corollary of the above, if $ \lambda $ and $ \omega $ are real, the ratio $ (\mathrm{i}V_{11}-\omega)/(\mathrm{i}V_{11}+\omega) $ is real and non-negative.
	\end{enumerate}
	We need an explanation for the latter part of (iv). If $ \mathrm{i}V_{11}+|\omega| $ changes its sign at some point $ x $, then the sign of $ \mathrm{i}V_{11}-|\omega| $ also must change simultaneously. However, it is impossible since $ \mathrm{i}V_{11} $ is continuous and  $ |\omega|>0 $. \\
	\indent From the above information (i)--(v), let us estimate the asymptotic behavior of Eqs.~(\ref{eq:solutemp}) and (\ref{eq:solvtemp}). When $ \lambda=-E $ is real, the integrand in the exp term $ \frac{U_{12}}{V_{12}}+\frac{U_{21}}{V_{21}}=2 \operatorname{Re}\frac{U_{12}}{V_{12}} $ is a real-valued function. Therefore, if $ \omega $ is real, the absolute value of the exponential term is equal to unity. Furthermore, using the fact (v) and Eq.~(\ref{eq:factorV12V21}), when $ \lambda $ and $ \omega $ are real, the relations 
	\begin{align}
		|u|^2=|C|\left|\mathrm{i}V_{11}-\omega\right|, 
\quad
		|v|^2=|C|\left|\mathrm{i}V_{11}+\omega\right|  \label{eq:absvforreal}
	\end{align}
	follow, because, e.g., 
	\begin{align}
		|V_{12}|\sqrt{\frac{\mathrm{i}V_{11}-\omega}{\mathrm{i}V_{11}+\omega}}=\sqrt{V_{12}V_{21}\frac{\mathrm{i}V_{11}-\omega}{\mathrm{i}V_{11}+\omega}}=\left|\mathrm{i}V_{11}-\omega\right|
	\end{align}
	holds. The expression (\ref{eq:absvforreal}) is obviously bounded if $ \Delta(x) $ is a bounded function. Thus, if $ \omega=\pm\sqrt{\det V} $ is real, the corresponding energy is in the spectrum.  On the other hand, if $ \lambda $ is real but $ \omega $ is pure imaginary, the exp term diverges exponentially. To sum up, 
	\begin{enumerate}[(A)]
		\item If $ \omega^2=\det V>0 $ for a given $ E $, this energy is in the spectrum.
		\item The band edges are determined by the equation $ \omega^2=\det V=0 $, which is an equation of order $ 2n+2 $ with respect to $ \lambda \ (\text{or }E) $ if the gap $ \Delta(x) $ is a solution of  $ \text{AKNS}_n $. So there exist $ 2n+2 $ band edges in general.
In this case, the solutions (\ref{eq:solutemp}) and (\ref{eq:solvtemp}) are drastically simplified:
	\begin{align}
	(u^2, v^2) = ( V_{12}, -V_{21}), \qquad (\omega=0). \label{eq:omega=0}
	\end{align}

		\item If $ \omega^2=\det V<0 $ for a given $ E $, this energy is not in the spectrum. The solution diverges exponentially and unphysical.
	\end{enumerate}
	\indent Since the discrete spectrum can be regarded as a shrinking limit of continuous spectrum, as a special case of (B) we  arrive at:
	\begin{enumerate}[(A)]
	\setcounter{enumi}{3}
		\item If the equation $ \det V=0 $ has a multiple root, this root belongs to the discrete spectrum.
The corresponding eigenstate can be a normalizable localized mode. 
	\end{enumerate}
	
Here we give a few technical remarks. If the gap function $ \Delta(x) $ is a solution of  $ \text{AKNS}_{n} $, it also can be a solution of higher order  $ \text{AKNS}_{m} \ (m \ge n) $. However, when we calculate the position of band edges, we should use the matrix $ V $ for the lowest order $ \text{AKNS}_n $. If we use a higher order one, we might obtain additional dummy edges and dummy discrete eigenvalues.\\
	\indent If $ \lambda $ and $ \omega $ are real, in other words, if $ E=-\lambda $ is in the spectrum, one can explicitly write down the square root of (\ref{eq:solutemp}) and (\ref{eq:solvtemp}) with the use 
of Eq.~(\ref{eq:absvforreal}). It is given by
	\begin{align}
	\begin{split}
		\begin{pmatrix}u \\ v \end{pmatrix}&=\begin{pmatrix}\sqrt{\left|\mathrm{i}V_{11}-\omega\right|}\mathrm{e}^{\frac{\mathrm{i}}{2}\arg V_{12}} \\ \mathrm{i}\operatorname{sgn}(\mathrm{i}V_{11})\sqrt{\left|\mathrm{i}V_{11}+\omega\right|}\mathrm{e}^{-\frac{\mathrm{i}}{2}\arg V_{12}}  \end{pmatrix} \\
		& \qquad\qquad\quad \times\exp\left[ \frac{\mathrm{i}\omega}{2}\int_0^x\mathrm{d}x\left( \frac{U_{12}}{V_{12}}+\frac{U_{21}}{V_{21}} \right) \right],
	\end{split}
	\end{align}
	where one should recall (iv), \textit{i.e.}, 
	\begin{align}
		\operatorname{sgn}(\mathrm{i}V_{11})=\operatorname{sgn}(\mathrm{i}V_{11}+\omega)=\operatorname{sgn}(\mathrm{i}V_{11}-\omega)
	\end{align} 
	is independent of $x$ and this sign factor is necessary to satisfy the relation (\ref{eq:uvratio}).

\section{Example: Complex Two-kink solution in  $ \text{AKNS}_2 $ }
	In this section, as an example, we consider the complex-valued two-kink state in  $ \text{AKNS}_2 $ and explicitly write down fermionic wave functions $ (u,v) $ in this state. Let  $ V $ be
	\begin{align}
		V = c_1 V^{(1)}+c_2 V^{(2)}+c_3 V^{(3)}+V^{(4)},
	\end{align}
	where we set $ c_4=1 $ without loss of generality. We then obtain the fundamental equation
	\begin{align}
		c_1 q-\mathrm{i}c_2q_x+c_3(-q_{xx}+2|q|^2q)+\mathrm{i}(q_{xxx}-6|q|^2q_x)=0
	\end{align}
	with setting $ r=q^* $. We can construct the following three independent integration constants by the method given in Ref. \cite{WadatiSanukiKonno}:
	\begin{align}
		J_i = \sum_{j=1}^4 J_{ij}c_j, \qquad (i=1,2,3),
	\end{align}
	where $ J_{ij} $ is a skew-symmetric  $ 4\times 4 $ matrix with components
	\begin{align}
		&J_{12}=rq,\qquad J_{13}=\mathrm{i}(r_xq-rq_x), \nonumber \\
		\begin{split}
		&J_{14}=3r^2q^2+r_xq_x-r_{xx}q-rq_{xx}, \\
		&J_{23}=-r^2q^2+r_xq_x,\qquad J_{24}=\mathrm{i}(r_{xx}q_x-r_xq_{xx}), 
		\end{split} \\
		&J_{34}=4r^3q^3+r^2q_x^2+r_x^2q^2-2rq(rq_{xx}+r_xq_x+r_{xx}q)+r_{xx}q_{xx} \nonumber
	\end{align}
	Using  $ J_1,\,J_2, $ and $ J_3 $, we can show the constancy of  $ \det V $: 
	\begin{align}
	\begin{split}
		\det V =& 16\lambda^6-16c_3\lambda^5+4(2c_2+c_3^2)\lambda^4-4(c_1+c_2c_3)\lambda^3\\
		&+(c_2^2+2c_1c_3-4J_1)\lambda^2+(2J_2+2c_3J_1-c_1c_2)\lambda \\
		&+\frac{1}{4}c_1^2-c_2J_1-c_3J_2-J_3.
	\end{split}
	\end{align}

\begin{figure}[t]
		\begin{center}
		\begin{tabular}{ll}
		\begin{minipage}{10.5em} (a)  $ \theta_1=\frac{\pi}{4},\ \theta_2=\frac{\pi}{4}\times1.01, $ \\ \hphantom{(a)}\! $ x_2-x_1=0 $.\end{minipage} & \begin{minipage}{10.5em} (d)  $ \theta_1=\frac{\pi}{4},\ \theta_2=\frac{\pi}{4}\times1.05, $ \\ \hphantom{(d)}\! $ x_2-x_1=0 $.\end{minipage} \\
		\includegraphics[scale=0.92]{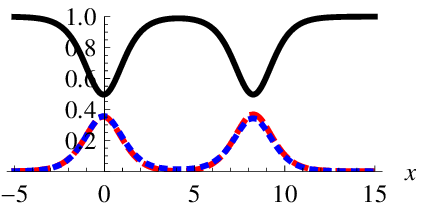} & \includegraphics[scale=0.92]{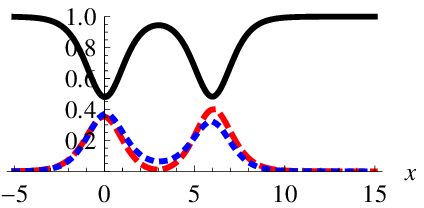} \\
		\begin{minipage}{10.5em} (b)  $ \theta_1=\frac{\pi}{4},\ \theta_2=\frac{\pi}{4}\times1.01, $ \\ \hphantom{(b)}\! $ x_2-x_1=2 $.\end{minipage} & \begin{minipage}{10.5em} (e)  $ \theta_1=\frac{\pi}{4},\ \theta_2=\frac{\pi}{4}\times1.012, $ \\ \hphantom{(e)}\! $ x_2-x_1=0.6 $.\end{minipage} \\
		\includegraphics[scale=0.92]{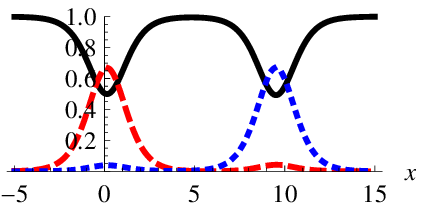} & \includegraphics[scale=0.92]{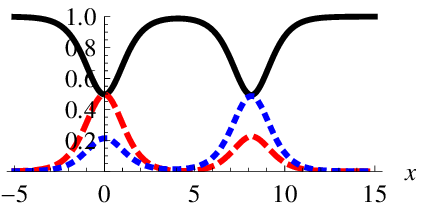} \\
		\begin{minipage}{10.5em} (c)  $ \theta_1=\frac{\pi}{4},\ \theta_2=\frac{\pi}{4}\times1.01, $ \\ \hphantom{(c)}\! $ x_2-x_1=4 $.\end{minipage} & \begin{minipage}{10.5em} (f)  $ \theta_1=\frac{\pi}{4},\ \theta_2=\frac{\pi}{4}\times1.015, $ \\ \hphantom{(f)}\! $ x_2-x_1=1.4 $.\end{minipage} \\
		\includegraphics[scale=0.92]{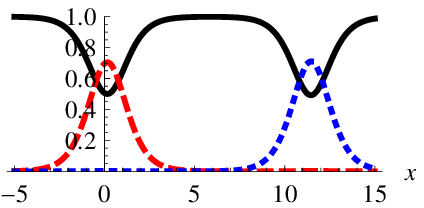} & \includegraphics[scale=0.92]{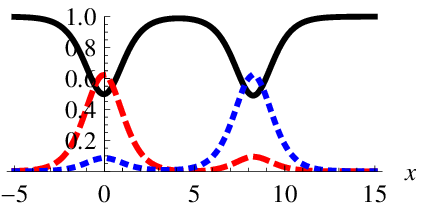}
		\end{tabular}
		\caption{\label{fig:complextwokinks}(Color online) Two-kink solution of the gap $|\Delta(x)|^2$ (black solid lines) and fermionic bound state wave functions, $2(|u_1(x)|^2+|v_1(x)|^2)$ (red dashed lines) and $2(|u_2(x)|^2+|v_2(x)|^2)$ (blue dotted lines), for various parameters (we set $ m=1 $). The bound states are delocalized around the two kinks when phase shifts of kinks are close to each other and $ x_2=x_1 $ [(a)]. As increasing $|x_2-x_1|$ with fixed $|\theta_1-€\theta_2|$, the distance between the kinks increases and consequently the bound states are localized on each kink [(a)$\rightarrow$(b)$\rightarrow$(c)]. As increasing $|\theta_1-€\theta_2|$ with fixed $|x_2-x_1|$, the distance between the kinks decreases while the delocalization character of the bound states is preserved [(a)$\rightarrow$(d)]. In (a)$\rightarrow$(e)$\rightarrow$(f), the kink profile is almost unchanged but the delocalization of bound states is lifted. Thus, the delocalization phenomenon of bound states is sensitive to the choice of kink-parameters.
}
		\end{center}
	\end{figure}

	As a special case, we consider the asymptotically uniform boundary condition
	\begin{align}
		\Delta = \mathrm{i}q\rightarrow\begin{cases} m&(x\rightarrow-\infty), \\ m\mathrm{e}^{-2\mathrm{i}\theta}&(x\rightarrow+\infty). \end{cases}
	\end{align}
	This boundary condition gives a complex-valued two-kink solution. In this case, we obtain
	\begin{align}
	\begin{split}
		c_1&=-2c_3m^2,\quad J_1=m^2c_2+3m^4, \\
		J_2&=m^4c_3,\quad J_3=m^4c_2+4m^6, \\
		\det V &= (\lambda^2-m^2)(4\lambda^2-2c_3\lambda+c_2+2m^2)^2.
	\end{split}
	\end{align}
	Introducing $ \theta_1 $ and $ \theta_2 $ by the relations
	\begin{align}
		\! c_2=2m^2(2\cos\theta_1\cos\theta_2\!-\!1),\ \;
		c_3=-2m(\cos\theta_1\!+\!\cos\theta_2),
	\end{align}
	with $\theta=\theta_1+\theta_2$, then $ \det V $ is factorized as
	\begin{align}
		\det V = 16(\lambda^2-m^2)(\lambda+m\cos\theta_1)^2(\lambda+m\cos\theta_2)^2.
	\end{align}
	Therefore the two-kink solution has a continuous spectrum for $ |E|\ge m $ and two bound states at $ E=m\cos\theta_1 $ and $ m\cos\theta_2 $. \\
	\indent The two-kink solution can be written as follows. Without loss of generality we assume  $ 0< \theta_j<\pi \ (j=1,2) $. Introducing the notations
	\begin{align}
	\begin{split}
		\kappa_1 &= m\sin\theta_1,\ \kappa_2 = m\sin\theta_2,\ \alpha = \frac{2\sqrt{\kappa_1\kappa_2}}{\mathrm{i}m(\mathrm{e}^{-\mathrm{i}\theta_1}-\mathrm{e}^{\mathrm{i}\theta_2})}, \\ 
		h_1(x) &= \frac{-\kappa_1(1+\mathrm{e}^{-2\kappa_2(x-x_2)})+\alpha\sqrt{\kappa_1\kappa_2}}{(1+\mathrm{e}^{-2\kappa_1(x-x_1)})(1+\mathrm{e}^{-2\kappa_2(x-x_2)})-|\alpha|^2}, \\ 
		h_2(x) &= \frac{-\kappa_2(1+\mathrm{e}^{-2\kappa_1(x-x_1)})+\alpha^*\sqrt{\kappa_1\kappa_2}}{(1+\mathrm{e}^{-2\kappa_1(x-x_1)})(1+\mathrm{e}^{-2\kappa_2(x-x_2)})-|\alpha|^2},
	\end{split}
	\end{align}
	with real constants $ x_1 $ and $ x_2 $ related to kink positions shown below, $ \Delta(x) $ can be expressed as\footnote{The definition of $ x_i \ (i=1,2) $ in this Letter differs from that in Ref. \cite{TakahashiNitta}. Here, we use  $ e_j(x)=\sqrt{\kappa_j}\mathrm{e}^{\kappa_j(x-x_j)} $ instead of $ e_j(x)=\mathrm{e}^{\kappa_j(x-x_j)} $ for \cite{TakahashiNitta}. }
	\begin{align}
		\Delta(x) = m+2\mathrm{i}(\mathrm{e}^{-\mathrm{i}\theta_1}h_1(x)+\mathrm{e}^{-\mathrm{i}\theta_2}h_2(x)). \label{eq:Ctwokinksol}
	\end{align}
	\indent The normalized bound state for $ E=m\cos\theta_j \ (j=1,2) $ is given by
	\begin{align}
		\begin{pmatrix}u_j(x) \\ v_j(x) \end{pmatrix}=\frac{\mathrm{e}^{-\kappa_i(x-x_j)}}{\sqrt{\kappa_j}}\begin{pmatrix}h_j(x) \\ \mathrm{e}^{\mathrm{i}\theta_j}h_j(x)^* \end{pmatrix},
	\end{align}
	and the scattering state with eigenenergy $ E $ is given by
	\begin{align}
		\begin{pmatrix}u(x,E) \\ v(x,E) \end{pmatrix} &= \mathrm{e}^{\mathrm{i}kx}\left[ \begin{pmatrix}m \\ E-k \end{pmatrix}+2\mathrm{i}\sum_{j=1,2}\frac{m}{m\mathrm{e}^{\mathrm{i}\theta_j}-E-k}\begin{pmatrix} h_j(x) \\ \mathrm{e}^{\mathrm{i}\theta_j}h_j(x)^* \end{pmatrix} \right], \nonumber \\
 k &=\pm\sqrt{E^2-m^2}.
	\end{align}
	We note that any two-kink solution can become self-consistent when the number of flavors is sufficiently large \cite{TakahashiNitta}. \\ 
	\indent The two-kink solution given by Eq.~(\ref{eq:Ctwokinksol}) has four parameters,  $ \theta_1, \theta_2, x_1, $ and $ x_2 $. If we exclude a trivial translational degree of freedom, there are actually three parameters, that is, $ \theta_1, \theta_2, $ and $ x_2-x_1 $. The two kinks are well separated from each other if the following condition holds
	\begin{align}
		m|x_2-x_1| \gg 1  \quad \text{or} \quad |\theta_1-\theta_2|\ll 1. \label{eq:separatedcondition}
	\end{align}
	In this case, an approximate position of each kink is given by
	\begin{align}
		x_{\text{left}} &\simeq \begin{cases} x_1 & (x_2\gg x_1) \\ -\frac{\log(\mathrm{e}^{-2\kappa x_1}+\mathrm{e}^{-2\kappa x_2})}{2\kappa}+\frac{(17m^2-8\kappa^2)\delta^2}{32\kappa^3\cosh^2\kappa(x_2-x_1)} & (x_2\simeq x_1),  \end{cases}  \\
		x_{\text{right}} &\simeq \begin{cases} x_2+\frac{1}{2\kappa_2}\log\frac{1-\cos\theta}{1-\cos\delta} & (x_2\gg x_1) \\ \frac{1}{2\kappa}\log\frac{1-\cos\theta}{1-\cos\delta}+\frac{\log(\mathrm{e}^{2\kappa x_1}+\mathrm{e}^{2\kappa x_2})}{2\kappa} & (x_2\simeq x_1). \end{cases} 
	\end{align}
	Here, we have defined 
$\delta=\theta_1-\theta_2$  and 
$ \kappa=m\sin(\theta/2) $. 
The gap function satisfies $\Delta(x) \sim m \mathrm{e}^{-2\mathrm{i}\theta_1}$ for $x_{\text{left}} \ll x \ll x_{\text{right}}$ 
and $\Delta(x) \sim m \mathrm{e}^{-2\mathrm{i}(\theta_1 +\theta_2)} = m \mathrm{e}^{-2\mathrm{i}\theta}$ 
for $x_{\text{right}} \ll x$ so that $-2\theta_i$ can be identified 
with the phase shift of the $i$-th kink.\\ 
\indent Figure \ref{fig:complextwokinks} shows the complex-two-kink solutions and their bound states for various cases.  
When the two kinks are well separated, the fermion bound states are localized 
around each kink in most parameter region.
When two kinks have phase shifts close to each other,
distance between them is shortest for $x_1=x_2$.
Only around this parameter region, {\it i.e.}, only when
	\begin{align}
		m|x_2-x_1| \lesssim 1 \quad  \text{and} \quad |\theta_1-\theta_2|\ll 1
	\end{align}
hold, the both fermion bound states have two peaks at the two kinks, 
that is, the ``delocalization'' of the bound states occurs.

%%%%%%%%%%%%%%%%%%%
%%%%%%%%%%%%%%%
\section{Summary and Discussion}
We have analytically derived the general expression 
for exact fermionic solutions 
(\ref{eq:solutemp}) and (\ref{eq:solvtemp}) 
to the BdG equation 
for all gap functions $\Delta(x)$ 
in the arbitrary order of the NLS hierarchy in the AKNS formalism, 
originally suggested by Correa, Dunne and Plyushchay 
\cite{Correa:2009xa}.
Depending on the sign of $\omega^2 = \det V$, 
these solutions are 
inside an energy band in the energy spectrum ($\det V > 0$), 
on a band edge ($\det V =0$), 
or exponentially divergent and unphysical ($\det V<0$).
The energy spectrum of our fermion solutions 
for the AKNS$_n$ contains 
two continuums for scattering states  
and $n$ bands for localized states. 
Consequently, there exist $2n+2$ band edges, 
on which the fermion solutions become rather simple 
as in Eq.~(\ref{eq:omega=0}). 
When $\det V=0 $ has a multiple root, 
the corresponding fermion solution belongs 
to the discrete spectrum. 

As an illustration of our formalism, 
we have presented the complex-two-kink solution 
for the AKNS$_2$ with the uniform boundary conditions for $\Delta(x)$ 
at $x \to \pm \infty$. 
The two kinks can be placed at any position and have phase shifts. 
In this example, the equation $\det V=0$  
has two multiple roots, 
giving two fermions localized on each or both of the two kinks. 
When the two kinks are well separated, the fermion bound states are localized 
around each kink in most parameter region.
When two kinks have phase shifts close to each other and are placed at distance as short as possible, the both fermion bound states have two peaks at the two kinks.
The self-consistency of the solutions 
is shown in \cite{TakahashiNitta} 
in which the $n$-kink solution is given.
While the uniform boundary condition is assumed for this example, 
which can be studied by the inverse scattering method, 
let us stress that 
our fermionic solutions can be applied to any solution 
under arbitrary boundary conditions. 
We may extend our formalism 
to unconventional superconductors such as 
multi-gap superconductors and $p$-wave superconductors.

\textbf{Acknowledgments}
We would like to thank G. Marmorini and T. Mizushima 
for useful discussions, and J. Feinberg and M. Thies for valuable comments. 
This work is supported in part by KAKENHI,  
Nos. 23740198 (MN), 23103515(MN) and 24740276 (ST).

%% References
%%
%% Following citation commands can be used in the body text:
%% Usage of \cite is as follows:
%%   \cite{key}          ==>>  [#]
%%   \cite[chap. 2]{key} ==>>  [#, chap. 2]
%%   \citet{key}         ==>>  Author [#]

%% References with bibTeX database:

\bibliographystyle{model1-num-names}
%\nocite{*}
%\bibliography{AKNSandBdG200}
%\bibliography{<your-bib-database>}

\begin{thebibliography}{33}
\expandafter\ifx\csname natexlab\endcsname\relax\def\natexlab#1{#1}\fi
\providecommand{\bibinfo}[2]{#2}
\ifx\xfnm\relax \def\xfnm[#1]{\unskip,\space#1}\fi
%Type = Book
\bibitem[{de~Gennes(1999)}]{DeGennes:1999}
\bibinfo{author}{P.~G. de~Gennes}, \bibinfo{title}{Superconductivity of metals
  and alloys}, \bibinfo{publisher}{Westview Press}, \bibinfo{year}{1999}.
%Type = Article
\bibitem[{Takayama et~al.(1980)Takayama, Lin-Liu, and Maki}]{Takayama:1980zz}
\bibinfo{author}{H.~Takayama}, \bibinfo{author}{Y.~R. Lin-Liu},
  \bibinfo{author}{K.~Maki},
\newblock \bibinfo{journal}{Phys. Rev. B} \bibinfo{volume}{21}
  (\bibinfo{year}{1980}) \bibinfo{pages}{2388}.
%Type = Article
\bibitem[{Brazovskii et~al.(1980)Brazovskii, Gordynin, and
  Kirova}]{Brazovskii1}
\bibinfo{author}{S.~A. Brazovskii}, \bibinfo{author}{S.~A. Gordynin},
  \bibinfo{author}{N.~N. Kirova},
\newblock \bibinfo{journal}{Pis. Zh. Eksp. Teor. Fiz.} \bibinfo{volume}{31}
  (\bibinfo{year}{1980}) \bibinfo{pages}{486}.
%Type = Article
\bibitem[{Brazovskii et~al.(1981)Brazovskii, Gordynin, and
  Kirova}]{Brazovskii2}
\bibinfo{author}{S.~A. Brazovskii}, \bibinfo{author}{S.~A. Gordynin},
  \bibinfo{author}{N.~N. Kirova},
\newblock \bibinfo{journal}{Pis. Zh. Eksp. Teor. Fiz.} \bibinfo{volume}{33}
  (\bibinfo{year}{1981}) \bibinfo{pages}{6}.
%Type = Article
\bibitem[{Brazovskii et~al.(1984)Brazovskii, Kirova, and
  Matveenko}]{Brazovskii3}
\bibinfo{author}{S.~A. Brazovskii}, \bibinfo{author}{N.~N. Kirova},
  \bibinfo{author}{S.~I. Matveenko},
\newblock \bibinfo{journal}{Zh. Eksp. Teor. Fiz.} \bibinfo{volume}{86}
  (\bibinfo{year}{1984}) \bibinfo{pages}{743}.
%Type = Article
\bibitem[{Heeger et~al.(1988)Heeger, Kivelson, Schrieffer, and
  Su}]{Heeger:1988zz}
\bibinfo{author}{A.~J. Heeger}, \bibinfo{author}{S.~Kivelson},
  \bibinfo{author}{J.~R. Schrieffer}, \bibinfo{author}{W.-P. Su},
\newblock \bibinfo{journal}{Rev. Mod. Phys.} \bibinfo{volume}{60}
  (\bibinfo{year}{1988}) \bibinfo{pages}{781}.
%Type = Article
\bibitem[{Machida and Fujita(1984)}]{Machida}
\bibinfo{author}{K.~Machida}, \bibinfo{author}{M.~Fujita},
\newblock \bibinfo{journal}{Phys. Rev. B} \bibinfo{volume}{30}
  (\bibinfo{year}{1984}) \bibinfo{pages}{5284}.
%Type = Article
\bibitem[{Fawcett(1988)}]{Fawcett}
\bibinfo{author}{E.~Fawcett},
\newblock \bibinfo{journal}{Rev. Mod. Phys.} \bibinfo{volume}{60}
  (\bibinfo{year}{1988}) \bibinfo{pages}{209}.
%Type = Article
\bibitem[{Jackiw and Rebbi(1976)}]{Jackiw:1975fn}
\bibinfo{author}{R.~Jackiw}, \bibinfo{author}{C.~Rebbi},
\newblock \bibinfo{journal}{Phys. Rev. D} \bibinfo{volume}{13}
  (\bibinfo{year}{1976}) \bibinfo{pages}{3398}.
%Type = Article
\bibitem[{Niemi and Semenoff(1986)}]{Niemi:1984vz}
\bibinfo{author}{A.~J. Niemi}, \bibinfo{author}{G.~W. Semenoff},
\newblock \bibinfo{journal}{Phys. Rept.} \bibinfo{volume}{135}
  (\bibinfo{year}{1986}) \bibinfo{pages}{99}.
%Type = Article
\bibitem[{Gross and Neveu(1974)}]{Gross:1974jv}
\bibinfo{author}{D.~J. Gross}, \bibinfo{author}{A.~Neveu},
\newblock \bibinfo{journal}{Phys. Rev. D} \bibinfo{volume}{10}
  (\bibinfo{year}{1974}) \bibinfo{pages}{3235}.
%Type = Article
\bibitem[{Nambu and Jona-Lasinio(1961)}]{Nambu:1961tp}
\bibinfo{author}{Y.~Nambu}, \bibinfo{author}{G.~Jona-Lasinio},
\newblock \bibinfo{journal}{Phys. Rev.} \bibinfo{volume}{122}
  (\bibinfo{year}{1961}) \bibinfo{pages}{345}.
%Type = Article
\bibitem[{Dashen et~al.(1975)Dashen, Hasslacher, and Neveu}]{Dashen:1975xh}
\bibinfo{author}{R.~F. Dashen}, \bibinfo{author}{B.~Hasslacher},
  \bibinfo{author}{A.~Neveu},
\newblock \bibinfo{journal}{Phys. Rev. D} \bibinfo{volume}{12}
  (\bibinfo{year}{1975}) \bibinfo{pages}{2443}.
%Type = Article
\bibitem[{Campbell and Bishop(1981)}]{Campbell:1981}
\bibinfo{author}{D.~K. Campbell}, \bibinfo{author}{A.~R. Bishop},
\newblock \bibinfo{journal}{Phys. Rev.} \bibinfo{volume}{B 24}
  (\bibinfo{year}{1981}) \bibinfo{pages}{4859--4862}.
%Type = Article
\bibitem[{Campbell and Bishop(1982)}]{Campbell:1981dc}
\bibinfo{author}{D.~K. Campbell}, \bibinfo{author}{A.~R. Bishop},
\newblock \bibinfo{journal}{Nucl. Phys. B} \bibinfo{volume}{200}
  (\bibinfo{year}{1982}) \bibinfo{pages}{297}.
%Type = Article
\bibitem[{Okuno and Onodera(1983)}]{OkunoOnodera}
\bibinfo{author}{S.~Okuno}, \bibinfo{author}{Y.~Onodera},
\newblock \bibinfo{journal}{J. Phys. Soc. Jpn.} \bibinfo{volume}{52}
  (\bibinfo{year}{1983}) \bibinfo{pages}{3495}.
%Type = Article
\bibitem[{Feinberg(2003)}]{FeinbergPLB}
\bibinfo{author}{J.~Feinberg},
\newblock \bibinfo{journal}{Phys. Lett. B} \bibinfo{volume}{569}
  (\bibinfo{year}{2003}) \bibinfo{pages}{204}.
%Type = Article
\bibitem[{Feinberg(2004)}]{Feinberg:2003qz}
\bibinfo{author}{J.~Feinberg},
\newblock \bibinfo{journal}{Annals Phys.} \bibinfo{volume}{309}
  (\bibinfo{year}{2004}) \bibinfo{pages}{166--231}.
%Type = Article
\bibitem[{Larkin and Ovchinnikov(1964)}]{larkin:1964zz}
\bibinfo{author}{A.~I. Larkin}, \bibinfo{author}{Y.~N. Ovchinnikov},
\newblock \bibinfo{journal}{Zh. Eksp. Teor. Fiz.} \bibinfo{volume}{47}
  (\bibinfo{year}{1964}) \bibinfo{pages}{1136}.
%Type = Article
\bibitem[{Machida and Nakanishi(1984)}]{Machida:1984zz}
\bibinfo{author}{K.~Machida}, \bibinfo{author}{H.~Nakanishi},
\newblock \bibinfo{journal}{Phys. Rev. B} \bibinfo{volume}{30}
  (\bibinfo{year}{1984}) \bibinfo{pages}{122}.
%Type = Article
\bibitem[{Shei(1976)}]{Shei:1976mn}
\bibinfo{author}{S.-S. Shei},
\newblock \bibinfo{journal}{Phys. Rev. D} \bibinfo{volume}{14}
  (\bibinfo{year}{1976}) \bibinfo{pages}{535}.
%Type = Article
\bibitem[{Ba{\c{s}}ar and Dunne(2008{\natexlab{a}})}]{Basar:2008im}
\bibinfo{author}{G.~Ba{\c{s}}ar}, \bibinfo{author}{G.~V. Dunne},
\newblock \bibinfo{journal}{Phys. Rev. Lett.} \bibinfo{volume}{100}
  (\bibinfo{year}{2008}{\natexlab{a}}) \bibinfo{pages}{200404}.
%Type = Article
\bibitem[{Ba{\c{s}}ar and Dunne(2008{\natexlab{b}})}]{Basar:2008ki}
\bibinfo{author}{G.~Ba{\c{s}}ar}, \bibinfo{author}{G.~V. Dunne},
\newblock \bibinfo{journal}{Phys. Rev. D} \bibinfo{volume}{78}
  (\bibinfo{year}{2008}{\natexlab{b}}) \bibinfo{pages}{065022}.
%Type = Article
\bibitem[{Fulde and Ferrell(1964)}]{Fulde:1964zz}
\bibinfo{author}{P.~Fulde}, \bibinfo{author}{R.~A. Ferrell},
\newblock \bibinfo{journal}{Phys.Rev.} \bibinfo{volume}{135}
  (\bibinfo{year}{1964}) \bibinfo{pages}{A550}.
%Type = Article
\bibitem[{Yoshii et~al.(2011)Yoshii, Tsuchiya, Marmorini, and
  Nitta}]{Yoshii:2011yt}
\bibinfo{author}{R.~Yoshii}, \bibinfo{author}{S.~Tsuchiya},
  \bibinfo{author}{G.~Marmorini}, \bibinfo{author}{M.~Nitta},
\newblock \bibinfo{journal}{Phys. Rev. B} \bibinfo{volume}{84}
  (\bibinfo{year}{2011}) \bibinfo{pages}{024503}.
%Type = Article
\bibitem[{Correa et~al.(2009)Correa, Dunne, and Plyushchay}]{Correa:2009xa}
\bibinfo{author}{F.~Correa}, \bibinfo{author}{G.~V. Dunne},
  \bibinfo{author}{M.~S. Plyushchay},
\newblock \bibinfo{journal}{Annals Phys.} \bibinfo{volume}{324}
  (\bibinfo{year}{2009}) \bibinfo{pages}{2522}.
%Type = Book
\bibitem[{Faddeev and Takhtajan(1987)}]{FaddeevTakhtajan}
\bibinfo{author}{L.~D. Faddeev}, \bibinfo{author}{L.~A. Takhtajan},
  \bibinfo{title}{Hamiltonian Methods in the Theory of Solitons},
  \bibinfo{publisher}{Springer}, \bibinfo{address}{Berlin, Heidelberg},
  \bibinfo{year}{1987}.
%Type = Article
\bibitem[{Hirota(1973)}]{Hirota1973}
\bibinfo{author}{R.~Hirota},
\newblock \bibinfo{journal}{J. Math. Phys.} \bibinfo{volume}{14}
  (\bibinfo{year}{1973}) \bibinfo{pages}{805}.
%Type = Article
\bibitem[{Sasa and Satsuma(1991)}]{SasaSatsumaJPSJ1991}
\bibinfo{author}{N.~Sasa}, \bibinfo{author}{J.~Satsuma},
\newblock \bibinfo{journal}{J. Phys. Soc. Jpn.} \bibinfo{volume}{60}
  (\bibinfo{year}{1991}) \bibinfo{pages}{409}.
%Type = Article
\bibitem[{Ablowitz et~al.(1974)Ablowitz, Kaup, Newell, and Segur}]{AKNS1974}
\bibinfo{author}{M.~J. Ablowitz}, \bibinfo{author}{D.~J. Kaup},
  \bibinfo{author}{A.~C. Newell}, \bibinfo{author}{H.~Segur},
\newblock \bibinfo{journal}{Stud. Appl. Math.} \bibinfo{volume}{53}
  (\bibinfo{year}{1974}) \bibinfo{pages}{249}.
%Type = Article
\bibitem[{Lax(1968)}]{Lax}
\bibinfo{author}{P.~D. Lax},
\newblock \bibinfo{journal}{Comm. Pure Appl. Math.} \bibinfo{volume}{21}
  (\bibinfo{year}{1968}) \bibinfo{pages}{467}.
%Type = Article
\bibitem[{Wadati et~al.(1975)Wadati, Sanuki, and Konno}]{WadatiSanukiKonno}
\bibinfo{author}{M.~Wadati}, \bibinfo{author}{H.~Sanuki},
  \bibinfo{author}{K.~Konno},
\newblock \bibinfo{journal}{Prog. Theor. Phys.} \bibinfo{volume}{53}
  (\bibinfo{year}{1975}) \bibinfo{pages}{419}.
%Type = Unpublished
\bibitem[{Takahashi and Nitta(6206)}]{TakahashiNitta}
\bibinfo{author}{D.~A. Takahashi}, \bibinfo{author}{M.~Nitta},
  \bibinfo{year}{arXiv:1209.6206}.

\end{thebibliography}

%% Authors are advised to submit their bibtex database files. They are
%% requested to list a bibtex style file in the manuscript if they do
%% not want to use model1-num-names.bst.

%% References without bibTeX database:

% \begin{thebibliography}{00}

%% \bibitem must have the following form:
%%   \bibitem{key}...
%%

% \bibitem{}

% \end{thebibliography}

\end{document}